\documentclass[12pt]{article}
\usepackage{amssymb,amsfonts}
\usepackage{graphics,psboxit,amsmath}
\usepackage{subfigure}
\usepackage{graphicx}
\usepackage{verbatim}
\usepackage[text={16.5cm,22.5cm},centering]{geometry}
\usepackage[all]{xy}

\usepackage{color}
\usepackage{hyperref}
\hypersetup{colorlinks}

\definecolor{darkred}{rgb}{1,0,0}
\definecolor{darkgreen}{rgb}{0,0.5,0}
\definecolor{darkblue}{rgb}{0,0,1}
\definecolor{orange}{rgb}{1,0.5,0}
\definecolor{green}{rgb}{0,1,0}
\definecolor{purple}{rgb}{.5,0,1}

\hypersetup{ colorlinks,
linkcolor=darkred,
filecolor=darkgreen,
urlcolor=darkblue,
citecolor=darkblue,
linktocpage=true }

\definecolor{markcolor}{rgb}{.25,0,1}

\definecolor{markcolor2}{rgb}{1,0,0}

\definecolor{markcolor3}{rgb}{0,1,0}

\newcommand{\no}{\noindent}

\def\be{\begin{equation}}
\def\ee{\end{equation}}
\def\ba{\begin{eqnarray}}
\def\ea{\end{eqnarray}}

\def\a{\alpha}

\def\d{\delta}

\def\e{\epsilon}

\def\m{\mu}
\def\n{\nu}
\def\om{\omega}

\def\l{\lambda}

\def\s{\sigma}

\begin{document}
\begin{titlepage}
\begin{center}
\strut\hfill
\vskip 1.3cm


\vskip .5in

{\Large \bf The $\frak{sl}({\cal N})$ twisted Yangian: bulk-boundary scattering $\&$ defects
}

\vskip 0.5in

{ \bf Jean Avan$^{a}$, Anastasia Doikou$^{b,c}$ and Nikos Karaiskos$^{d}$} \vskip 0.2in

{\footnotesize $^{a}$ Laboratoire de Physique Th\'eorique et Mod\'elisation
(CNRS UMR 8089), \\
Universit\'e de Cergy-Pontoise, F-95302 Cergy-Pontoise, France}
\\[4mm]
\noindent
{\footnotesize  $^b$Department of Mathematics, Heriot-Watt University,
\\ EH14 4AS, Edinburgh, United Kingdom}
\\[4mm]
\noindent
{\footnotesize $^{c}$ Department of Computer Engineering \& Informatics, \\
University of Patras, GR-Patras 26500, Greece}
\\[4mm]
\noindent
{\footnotesize  $^d$Institute for Theoretical Physics, Leibniz University
Hannover,\\ Appelstra\ss e 2, 30167 Hannover, Germany}

\vskip .1in


{\footnotesize {\tt E-mail: avan@u-cergy.fr, a.doikou@hw.ac.uk,
nikolaos.karaiskos@itp.uni-hannover.de}}\\

\end{center}

\vskip 1.0in

\centerline{\bf Abstract}
We consider the $\mathfrak{sl}({\cal N})$ twisted Yangian quantum spin chain. In particular, we study
the bulk and boundary scattering of the model via the solution of the Bethe ansatz equations in
the thermodynamic limit. Local defects are also implemented in the model and the associated
transmission amplitudes are derived through the relevant Bethe ansatz equations.
\no

\vfill

\end{titlepage}
\vfill \eject
\tableofcontents

\section{Introduction}

Spin chains can be considered as main paradigms of quantum integrable systems:
the discrete structure on the lattice allows to disregard complications of infinite
quantities arising from same-site manipulations of  generators, and the characteristic structure
of co-product (or alternatively co-module in the reflection case) of the underlying quantum algebras
(Yangian, Twisted Yangian, Quantum Groups, Reflection algebras, Elliptic Algebras)
directly translates into the site-by-site building of the space of quantum states and the Hamiltonians deduced from monodromy matrices.

Open spin chains, requiring introduction of boundary terms consistent with quantum integrability, are in particular related to generalized reflection algebras (quadratic algebras) \`a la Freidel-Maillet \cite{fm} extending the original construction of Cherednik \cite{cherednik} and Sklyanin \cite{sklyanin} to a four matrix structure canonically expressed as:
\be
A_{12}\ K_1\ B_{12}\ K_2 = K_2\ C_{12}\ K_1\ D_{12} \, ,
\label{quadr}
\ee
with unitarity requirements
\ba
&& A_{12}\ A_{21} = D_{12}\ D_{21} = {\mathbb I}_{12} \, ,\cr
&& C_{12}= B_{21}. \label{one}
\ea
In the particular case when $A_{12} =D_{21} =R_{12}$ a given Yang-Baxter $R$ matrix, and $B_{12} =C_{21} = \bar R_{21}$ (its soliton anti-soliton counter part), $\bar R_{12}\sim R_{12}^{t_1}$, (\ref{quadr}) yields the so-called twisted Yangian structure if $R$ is the simple Yangian solution of the Yang-Baxter equation \cite{Ols}.

Spin chains based on such a twisted Yangian were first constructed and investigated in \cite{Doikou:2000yw}. They were then
considered in the thermodynamic limit in our previous paper \cite{Avan:2014noa}. They naturally exhibit soliton non-preserving boundary conditions due the choice of $B_{12} = C_{21}$
as a soliton$-$anti-soliton $S$-matrix and the subsequent conversion of a soliton into an anti-soliton by the building reflection matrix $K$.

We describe here the additional effect of a defect in such a spin chain for a general $A_n$ underlying Lie algebra. Assuming that the spin chain is as usual derived from
a monodromy matrix obtained from the canonical representation of the co-module algebra by the coefficient matrices $A,C$ or $D,B$, defects can be naturally implemented in the context of spin chain model building by inserting a different representation of the co-module algebra into the double row transfer matrix at one physically meaningful site (the defect site): since we are dealing with open spin chains the defect is in fact inserted at two ``mirror'' imaged sites when building the monodromy matrix. Note that we consider here the case of a purely transmitting defect, the reflecting-transmitting defects require more intricate manipulations (see e.g. \cite{ragoucy}).

The paper runs as follows: we recall the Bethe ansatz equations (BAE) for the $\mathfrak{sl}({\cal N})$ twisted Yangian spin chain and define the energy of particle-like excitations. We then derive formulas for both bulk and boundary scattering amplitudes. We finally address the issue of computing the particle-defect transmission amplitudes. The corresponding quantization condition is also derived.

\section{BAE for $\frak{sl}({\cal N})$ twisted Yangian}

The main aim of this section is the study of the Bethe ansatz
equations in the thermodynamic limit. In particular, the ground
state and the low lying excitations of the model are identified.
The BAE for the $\mathfrak{sl}(\cal N)$ twisted Yangian were derived
in \cite{Arnaudon:2004sd}. We shall distinguish below two cases
${\cal N} = 2n$ or ${\cal N} = 2n+1$. Note that throughout the text we consider the boundary matrices, $c$-number representations of the twisted Yagian (\ref{one}) ($A_{12} =D_{21} =R_{12}$, and $B_{12} =C_{21} = \bar R_{21}$ ), to be proportional to unit.
Defining
\be
e_n(\l) = \frac{\l+ \frac{in}{2}}{\l - \frac{in}{2}} \, ,
\ee
the BAE read as follows:
\begin{itemize}
\item  $\mathfrak{sl}(2n+1)$
\be
\begin{split}
e^L_{1}(\l_i^{(1)}) & = -
\prod_{j=1}^{M^{(1)}} e_{2}(\l_i^{(1)} - \l_j^{(1)})\,
e_{2}(\l_{i}^{(1)} + \l_j^{(1)})
\prod_{j=1}^{M^{(2)}}
e_{-1}(\l_i^{(1)} - \l_j^{(2)}) \,
e_{-1}(\l_i^{(1)} + \l_j^{(2)})\,, \cr
1 & = - \prod_{j=1}^{M^{(\ell)}}
e_{2}(\l_i^{(\ell)} - \l_j^{(\ell)}) \,
e_{2}(\l_i^{(\ell)} + \l_j^{(\ell)})
\prod_{ \tau = \pm1}
\prod_{ j=1}^{M^{(\ell+\tau)}} e_{-1}(\l_i^{(\ell)} - \l_j^{(\ell+\tau)}) \,
e_{-1}(\l_{i}^{(\ell)} + \l_{j}^{(\ell+\tau)}) \cr
& \hspace{0cm} \textrm{for} \qquad \ell = 2, \ldots, n-1, \cr
e_{-\frac{1}{2}}(\l_i^{(n)}) & = - \prod_{j=1}^{M^{(n)}}
e_{-1}(\l_i^{(n)} - \l_j^{(n)})\,
e_{-1}(\l_i^{(n)} + \l_j^{(n)}) \,
e_2(\l_i^{(n)} - \l_j^{(n)}) \,
e_2(\l_i^{(n)} + \l_j^{(n)}) \cr
& \qquad  \times  \prod_{ j=1}^{M^{(n-1)}}
e_{-1}(\l_i^{(n)} - \l_j^{(n-1)}) \,
e_{-1}(\l_i^{(n)} + \l_j^{(n-1)})  \, .
\end{split}
\label{BAE1b}
\ee
Note that in this case the Bethe ansatz equations are similar to the ones of the open $\mathfrak{osp}(1|2n)$ spin chain (see also \cite{Doikou:2000yw}, \cite{Arnaudon:2004sd}, \cite{Arnaudon:2003zw}).
\item $\mathfrak{sl}(2n)$
\begin{eqnarray}
  e_{1}^L(\l_{i}^{(1)}) &\!\!=\!\!& -\prod_{j=1}^{M^{(1)}}
  e_{2}(\l_{i}^{(1)} - \l_{j}^{(1)})\ e_{2}(\l_{i}^{(1)} +
  \l_{j}^{(1)})\ \prod_{ j=1}^{M^{(2)}}e_{-1}(\l_{i}^{(1)} -
  \l_{j}^{(2)})\ e_{-1}(\l_{i}^{(1)} + \l_{j}^{(2)})\,,
  \nonumber \\
  1 &\!\!=\!\!& - \prod_{j=1}^{M^{(\ell)}} e_{2}(\l_{i}^{(\ell)} -
  \l_{j}^{(\ell)})\ e_{2}(\l_{i}^{(\ell)} + \l_{j}^{(\ell)})\
  \prod_{\tau = \pm 1}\prod_{ j=1}^{M^{(\ell+\tau)}}
  e_{-1}(\l_{i}^{(\ell)} -
  \l_{j}^{(\ell+\tau)})\ e_{-1}(\l_{i}^{(\ell)} +
  \l_{j}^{(\ell+\tau)}) \nonumber \\
  && \ell= 2,\ldots,n-1, \nonumber \\
  e_{-1}(\l_{i}^{(n)}) &\!\!=\!\!& - \prod_{j=1}^{M^{(n)}}
  e_{2}(\l_{i}^{(n)} - \l_{j}^{(n)})\ e_{2}(\l_{i}^{(n)} +
  \l_{j}^{(n)}) \nonumber \\
  && \times \prod_{ j=1}^{M^{(n-1)}}e_{-1}^{2}(\l_{i}^{(n)} -
  \l_{j}^{(n-1)})\ e_{-1}^{2}(\l_{i}^{(n)} + \l_{j}^{(n-1)}).
\label{BAE2}
\end{eqnarray}
As opposed to the $\mathfrak{sl}(2n+1)$ case the Bethe ansatz equations above do not reduce to any of the known forms of BAE, which makes the whole study even more intriguing.

Note that the number $M^{(l)}$ are associated to the eigenvalues of the diagonal generators $S_l$ of the underlying algebra $\mathfrak{so}(n)$ (see \cite{Doikou:2000yw, Arnaudon:2004sd} for a detailed discussion on the underlying symmetry of the models), i.e.
\be
S_1 = {1\over 2} M^{(0)} - M^{(1)}, ~~~~S_l = M^{(l-1)} - M^{(l)}, ~~~~S_l = {1\over 2}(E_{ll} -E_{\bar l \bar l}), ~~~~ l\leq l \leq {{\cal N} -1 \over 2} \label{numbers}
\ee
$E_{ll}$ the diagonal generators of $\mathfrak{sl}({\cal N})$, and $\bar l = {\cal N} - l +1$ the conjugate index.

It is also worth recalling that the corresponding numbers in he usual $\mathfrak{sl}({\cal N})$ case are givan by:
\be
E_{ll} = M^{(l-1)} -M^{(l)}, ~~~~M^{(0)} = 2L, ~~~~~M^{({\cal N})}=0,  ~~~~l \in \{1, 2, \ldots, {\cal N}\}
\ee
By imposing $M^{(l)} = M^{({\cal N} -1)}$ and considering the differences $E_{ll} - E_{\bar l \bar l}$  we end up to (\ref{numbers}) in accordance to the folding of $\mathfrak{sl}({\cal N})$ leading to the $\mathfrak{so}(n)$ algebra \cite{Doikou:2000yw, Arnaudon:2004sd}.
\end{itemize}

The ground state of the model consists of $n$ filled Dirac seas, unlike
the Yangian case, where the ground state consists of $2n+1$ or $2n$ filled
seas respectively.
The number of seas is halved here due to the underlying algebraic
folding (see also \cite{Arnaudon:2004sd}).
As usual, an excitation corresponds to a hole in the Dirac sea. We perform
our computations in the thermodynamic limit of the BAE, which is obtained
according to the thermodynamic rule (for more details the interested reader in referred to e.g. \cite{Andrei:1983cb, Grisaru:1994ru} or \cite{Avan:2014noa} in a more relevant context)
\be
\frac{1}{L} \sum_{j=1}^{M^{(\ell)}} f(\l_j^{(\ell)}) \to
\int_0^{\infty} d\m \, \s_{\ell}(\m) \, f(\m) -
\frac{1}{L} \sum_{j=1}^{\n^{(\ell)}} f(\tilde{\l}_j^{(\ell)})
- \frac{1}{2L} f(0)\, ,
\ee
with $\n^{(\ell)}$ holes of rapidities $\tilde{\l}_j^{(\ell)}$ in the
$\ell^{\textrm{th}}$ Dirac sea $\s_{\ell}$ is the density in the $\ell^{th}$ sea. The last term is the halved contribution
at $0^+$ due to the boundaries. In the thermodynamic limit the BAE take the
compact form
\be
\begin{split}
 \hat{\mathcal{K}}(\om) \, \hat{\s}(\om) & =
\hat{\mathcal{A}}(\om) + \frac{1}{L}\Big (\hat{\mathcal{F}}^{(1)}(\om)
+ {\mathcal F}^{(2)}(\om)\Big ) \, \cr
\Rightarrow  \hat \s(\omega) & = \hat \s^{(0)}(\omega)
+ {1\over L} \Big (r^{(1)}(\omega) + r^{(2)}(\omega)\Big ) \, ,
\end{split}
\ee
where we have defined the $n$-dimensional column
vectors $\hat{\s}, \hat{\mathcal{A}}$ and
$\hat{\mathcal{F}}^{(i)}$ with elements
\be
\hat{\s}_{i}(\om)\qquad
\hat{\mathcal{A}}_{i}(\om) = \hat{a}_1(\om) \, \d_{i1} \, , ~~~~~~i\in \{1,\ 2, \ldots, n\} \, ,
\ee
also we define
\be
a_n(\lambda) = {i \over 2\pi} {d\over d\lambda}(\ln e_n(\lambda)), ~~~~~\hat a_n(\omega) = e^{-{n|\omega| \over 2}}
\ee
and
\begin{itemize}
\item  $\mathfrak{sl}(2n+1)$
\ba
&& \hat{\mathcal{F}}^{(1)}_i(\om)  = \hat{a}_1(\om) \d_{i1}
- 2\hat{a}_1(\om) + \hat{a}_2(\om)- \hat{a}_{\frac{1}{2}}(\om) \d_{i n}
\label{ff} \\
&& \hat{\mathcal{F}}^{(2)}_i(\om) =
 2 \Big(\hat{a}_2(\om) -\hat{a}_1(\om) \, \d_{i n}\Big)
\sum_{j=1}^{\n^{(i)}}
\cos\big(\om \tilde{\l}_j^{(i)} \big)- 2\hat{a}_1(\om) \sum_{j=1}^{\n^{(k)}}
\cos\big(\om \tilde{\l}_j^{(k)} \big) \big(
\d_{k,i+1} + \d_{k,i-1}
\big) \, . \nonumber
\ea
The kernel $\hat{\mathcal{K}}(\om)$ is the $n\times n$ matrix
\be
\hat{\mathcal{K}}_{ij}(\om) =
\big(1 - \hat{a}_1(\om) \d_{in} + \hat{a}_2(\om)\big)\d_{ij}
- \hat{a}_1(\om) \big(\d_{i,j+1} + \d_{i,j-1}\big) \, , \quad
i,\ j \in \{ 1,\cdots,n \}\, ,
\ee
and the entries of its inverse are given by
\be
\hat{\mathcal{R}}_{ij}(\om) =
e^{\frac{\om}{2}}
\frac{\sinh\big( \min(i,j) \frac{\om}{2}\big)
\cosh \big( n + \frac{1}{2} - \max(i,j)\big) \frac{\om}{2}}
{\cosh(n+\frac{1}{2})\frac{\om}{2} \,
\sinh \frac{\om}{2}} \, .
\ee
\item $\mathfrak{sl}(2n)$

In this case as well the ground state consists of $n$ filled Dirac seas. The
thermodynamic limit of the BAE leads to the densities of Bethe roots
(see expressions (\ref{ff}))
with $\hat{\mathcal{F}}^{(i)}$ now defined accordingly as
\ba
&& \hat{\mathcal{F}}^{(1)}_i(\om)  =
\hat{a}_2(\om) - (2 - \d_{i1} + \d_{in}) \, \hat{a}_1(\om)
 \\
&& \hat{\mathcal{F}}^{(2)}_i(\om)  =
- 2\hat{a}_1(\om) \sum_{j=1}^{\n^{(k)}}
\cos \big(\om \tilde{\l}_j^{(k)} \big)   \big(
\d_{k,i+1} + \d_{k,i-1} (1+\d_{in})
\big) + 2\hat{a}_2(\om) \sum_{j=1}^{\n^{(i)}}
\cos \big(\om \tilde{\l}_j^{(i)} \big)  \, . \nonumber
\ea
The kernel $\hat{\mathcal{K}}$ is given by the
$n \times n$ matrix with elements
\be
\mathcal{K}_{ij}(\om) = \big(1+\hat{a}_2(\om)\big) \d_{ij}
- \hat{a}_1(\om) (\d_{i,j+1} + \d_{i,j-1})
- \hat{a}_1(\om) \d_{in}\d_{j,n-1} \, ,
\ee
and its inverse by
\be
\hat{\mathcal{R}}_{ij}(\om) = e^{\frac{\om}{2}}
\frac{\sinh\big(\min(i,j)\frac{\om}{2}\big)
\cosh\big((n-\max(i,j))\frac{\om}{2}\big)}
{(1+\d_{jn})\cosh (\frac{n\om}{2}) \sinh (\frac{\om}{2})} \, .
\ee
\end{itemize}
Having set all the necessary ingredients we are now in a position to proceed
with the thermodynamic computations of the energy of particle-like
excitations, as well as their scattering amplitudes. Later in the text local
defects will be introduced and the scattering between the excitation
and the defects will be discussed in detail.

\subsection{The energy}

In this section the energy of particle-like excitations (holes in the
$\ell^{th}$ sea) is derived. From the eigenvalues of the transfer matrix,
the energy of the system may be derived taking the first derivative of the
transfer matrix eigenvalues with respect to the spectral parameter
(see also e.g. \cite{Avan:2014noa} for a more detailed computation).

As a validity check on the form of the ground state and excitations,
together with the quantization condition derived in \cite{Avan:2014noa}
(see also next section), we compute the energy of a single hole
in the $j^{th}$ sea and compare with the resulting expression for
the density $\hat{\s}_j^{(0)}$. These quantities should be the same so that the
quantization condition may be appropriately employed. It is worth noting that the a single excitations here is associated
to representations of the underlying exact symmetry which is $\mathfrak{so}(n)$ it is indeed a ``folded'' algebra (a folding at the level of Dynkin diagram occurs) compared to $\mathfrak{sl}(2n),\ \mathfrak{sl}(2n+1)$ as extensively discussed in \cite{Doikou:2000yw, Arnaudon:2004sd}, see also as similar discussion and examples on the quantum numbers of excitations in \cite{Avan:2014noa}.

The energy derived in the thermodynamic is given by the following
expression (see also \cite{Avan:2014noa}):
\be
\e = - \int_0^{\infty} d\m \, a_1(\m) \, \s_{1}(\m)
+ \frac{1}{L} a_1(\tilde{\l}_1) - \frac{1}{2L}a_1(0) \, .
\ee
The energy of a particle-like excitation (hole) in the $j^{th}$ sea
in particular is then given as (its Fourier transform)
\be
\hat \epsilon^{(j)}(\omega) = - \hat a_1(\omega) \hat r_j^{(2)}(\omega)
+\hat  a_1(\omega)\delta_{j1} \, ,
\ee
whereas the density $\hat \s^{(0)}_j$ is given by
\be
\hat \s^{(0)}_j = \hat {\cal R}_{j1}(\om)\ \hat a_1(\om).
\ee
We define the $\hat r_j^{(2)}(\omega)$ quantities as follows:
\begin{itemize}
\item $\mathfrak{sl}(2n+1)$
\ba
\hat r_1^{(2)}(\omega) &= &\hat R_{11}(\omega) \hat a_2(\omega) - \hat a_1(\omega)
\hat R_{12}(\omega), ~~\mbox{hole in the $1^{st}$ sea} \cr
\hat r_j^{(2)}(\omega) &= &\hat R_{1j}(\omega) \hat a_2(\omega) - \hat a_1(\omega)
(\hat R_{1j+1}(\omega)+\hat R_{1j-1}(\omega)), ~~\mbox{hole in the $j^{th}$ sea ($j \neq 1,\ n$)} \cr
\hat r_n^{(2)}(\omega) &= &\hat R_{1n}(\omega)( \hat a_2(\omega) - \hat a_1(\omega))
- \hat a_1(\omega) \hat R_{1n-1}(\omega), ~~\mbox{hole in the $n^{th}$ sea} \, .
\ea
It turns out that the energy of a hole in the $j^{th}$ sea is given as:
\be
\hat \epsilon^{(j)}(\omega) = {\cosh(n+{1\over 2} -j ){\om \over 2} \over \cosh(n +{1\over 2}){\om \over 2} },
~~~~j\in\{1, 2, \ldots , n\}
\ee
whereas the density $\hat{\s}^{(0)}_j$ is computed to be:
\be
\hat{\s}_j^{(0)}(\om) = \hat{\mathcal{R}}^j_{~1}(\om) \, \hat{a}_1(\om)
= \frac{\cosh(n+\frac{1}{2}-j)\frac{\om}{2}}{\cosh(n+\frac{1}{2})
\frac{\om}{2}} \, , \quad j \in\{ 1, \ldots , n \}\, .
\ee
\item
$\mathfrak{sl}(2n)$
\\
The expression of the energy of a hole in the $j^{th}$ sea is given by the same
expressions as in the odd case with the exception of the holes in the $n$ and $n-1$ seas:
\ba
\hat r_{n-1}^{(2)}(\omega) &= & \hat R_{1n-1}(\omega) \hat a_2(\omega) - \hat a_1(\omega)
(2\hat R_{1n}(\omega)+\hat R_{1n-2}(\omega)), ~~\mbox{hole in the $(n-1)^{th}$ sea ($j \neq 1,\ n$)} \cr
\hat r_n^{(2)}(\omega) &= &\hat R_{1n}(\omega) \hat a_2(\omega)  -
\hat a_1(\omega) \hat R_{1n-1}(\omega), ~~\mbox{hole in the $n^{th}$ sea} \, .
\ea
In this case the energy of a hole in the $j^{th}$ sea is given as:
\ba
&& \hat \epsilon^{(j)}(\om) = {\cosh(n-j){\omega \over 2} \over \cosh {n \omega \over 2}},
~~~~j\in\{1, 2, \ldots n-1\}\, , \cr
&& \hat \epsilon^{(n)}(\om) = {1 \over 2\cosh {n \omega \over 2}} \, ,
\ea
and the density $\hat \s^{(0)}_j$ is given by:
\ba
&& \hat \s_j^{(0)}(\om) = \hat{\mathcal{R}}_{j1} (\om) \, \hat{a}_1(\om)
= \frac{\cosh(n-j)\frac{\om}{2}}{\cosh \frac{n\om}{2}} \,,~~~~j\in\{1, \ldots n-1\} \, .\cr
&&
\ea
\end{itemize}
Having verified the fact that for each particle-like excitation the equation:
$\hat \s_j^{(0)}(\om) = \hat \epsilon^{(j)}(\om)$,
is valid and compatible with the
quantization condition \cite{Avan:2014noa}, we now derive the associated
bulk and boundary scattering amplitudes.

\section{Scattering amplitudes}

The key element in this context is now the generalized quantization
condition for the twisted Yangian introduced in \cite{Avan:2014noa}. We shall
consider here the scattering of particle-like excitations in the first sea.
Recall that the quantization condition (see also \cite{Korepin:1979hg},
\cite{Andrei:1983cb}, \cite{Grisaru:1994ru}) for a state with two holes in the
$\ell^{th}$ sea reads as \cite{Avan:2014noa}
\be
\Big( e^{i\mathcal{P}^{(\ell)}L} \,
{\mathbb S}(\tilde{\l}_1^{(\ell)}, \tilde{\l}_2^{(\ell)}) - 1
\Big)
|\tilde{\l}_1^{(\ell)} \tilde{\l}_2^{(\ell)} \rangle = 0 \, ,
\ee
where ${\cal P}^{(\ell)}$ is the momentum of the hole in the $\ell^{th}$ sea, and the global scattering matrix ${\mathbb S}$ is given by \cite{Avan:2014noa}
\be
{\mathbb S}(\lambda_1,\ \lambda_2) =  {\cal K}^+(\lambda_1)\ {\cal S}(\lambda_1-\lambda_2)\
{\cal K}^-(\lambda_2)\ {\cal S}(\lambda_1+\lambda_2) \, ,
\ee
where ${\cal S}(\l) = S(\l)\, \bar S(\l)$ and $S$ ($\bar{S}$) corresponds to the
soliton$-$soliton (soliton$-$anti-soliton) scattering amplitude of the Yangian
$\mathfrak{sl}({\cal N})$. This bulk factorization will be explicitly shown below.
${\cal K}^{\pm}$ are the physical boundary scattering matrices
associated with the left/right boundaries of the system. We have considered here
for simplicity the left/right boundary matrices ${\mathbb K}^{\pm} \propto {\mathbb I}$.
Note that here we obtain the eigenvalue associated to the hole-hole and hole-boundary interactions. Note that more eigenvalues can be identified using complex ``string'' type solutions of the BAE, however such an analysis is beyond the intended scope of the present article, given especially the algebraic arguments leading to the factorized structure of the scattering matrix (see also \cite{Avan:2014noa}).

Indeed the validity of the factorization (and the quantization condition for that matter) at the matrix level as well as the form of the  $S$ and $\bar S$ matrices are confirmed by the underlying algebra as well as the quantization condition (for similar algebraic arguments see \cite{Doikou-trans}, \cite{Avan:2014noa}). It is in any case well established that the Bethe asnatz formulation serves as a ``renormalization'' process, thus $S$, $\bar S$ and ${\cal K}$ matrices are basically ``renormalized'' (physical) quantities as opposed to the ``bare'' $R, \bar R$ and ${\mathbb K}^{\pm}$. Bethe ansatz  provides essentially the overall physical factors $S_0,\ \bar S_0$ and $K_0^{\pm}$, and this is exactly what we perform in what follows.

Using the dispersion relation
\be
\e^{(\ell)} (\l) = \frac{1}{2\pi} \frac{d}{d\l} \mathcal{P}^{(\ell)}(\l) \, ,
\ee
and the fact that $L\int_0^{\tilde{\l}_i^{(\ell)}} d\l \, \s(\l) \in \mathbb{Z}$,
we conclude that the scattering matrix phase, ($\mathbb{S}=\exp(i \bf{\Phi})$), is given by

\be
i{\bf \Phi} = -\int_{-\infty}^{\infty}
\frac{d\om}{\om} \, e^{-i\om \tilde{\l}_1}
\big(
\hat{\s}_{1}(\om) - \hat{\e}^{(1)}(\om)
\big) \, .
\ee
Introducing two excitations (holes) in the first sea gives
\be
i{\bf \Phi} = - \int_{-\infty}^{\infty}
\frac{d\om}{\om} \,e^{-i\om \tilde{\l}_1}
\, \sum_{j=1}^n \hat{\mathcal{R}}_{1j}(\om) \, \hat{\mathcal{F}}_j(\om) \, .
\ee
Recalling the expression for $\hat{\mathcal{F}}^{(i)}$, and keeping in mind that
we consider two holes in the first sea, we need the following quantities
in the summation above:
\be
\begin{split}
 \hat{\mathcal{F}}^{(1)}_1(\om) & =
 - \hat{a}_1(\om) + \hat{a}_2(\om)\, , ~~~~~ \hat{\mathcal{F}}^{(2)}_1(\om)= \hat{a}_2(\om)
 \sum_{j=1}^2 \big(
 e^{i\om\tilde{\l}_j^{(1)}} + e^{-i\om\tilde{\l}_j^{(1)}}
 \big) \cr
 \hat{\mathcal{F}}^{(1)}_2(\om) & = -2\hat{a}_1(\om) + \hat{a}_2(\om)\, , ~~~~
 \hat{\mathcal{F}}^{(2)}_2(\om) = - \hat{a}_1(\om) \sum_{j=1}^2 \big(
 e^{i\om\tilde{\l}_j^{(1)}} + e^{-i\om\tilde{\l}_j^{(1)}}
 \big) \cr
 \hat{\mathcal{F}}^{(1)}_k(\om) & = -2\hat{a}_1(\om) + \hat{a}_2(\om) \, ,
 \qquad  \hat{\mathcal{F}}^{(2)}_k(\om)=0 \, , \qquad k = 3, \ldots , n - 1 \, ,\cr
 \hat{\mathcal{F}}^{(1)}_n(\om) &  = \Big\{
 \begin{array}{ll}
  -2\hat{a}_1(\om) + \hat{a}_2(\om) -
 \hat{a}_{\frac{1}{2}}(\om)\, , & \frak{sl}(2n+1) \cr
  -3\hat{a}_1(\om) + \hat{a}_2(\om) \, , & \frak{sl}(2n)
 \end{array} \, ,
 \, ~~~~\hat{\mathcal{F}}^{(2)}_n(\om) =0.
\end{split}
\ee
We identify then the bulk and boundary scattering amplitudes as
\be
\begin{split}
 \mathcal{S}_0(\l) & = \exp\Big \{-\int_{-\infty}^\infty \frac{d\om}{\om} \,
 e^{-i\om {\l}} \, \mathcal{B}_1(\om)\Big \} \cr
 K_0^+(\l) \, K_0^-(\l) & = \exp\Big \{- \int_{-\infty}^{\infty}
 \frac{d\om}{\om} \, \Big(
 e^{-i\om \tilde{\l}^{(1)}} \, \mathcal{B}_2(\om)
 + e^{-2i\om {\l}} \, \mathcal{B}_1(\om)
 \Big)\Big \} \, ,
\end{split} \label{scatter}
\ee
where the terms $\mathcal{B}_i$ contain the collected contributions from
$\mathcal{F}^{(j)}$ and $\mathcal{R}$ and read
\be
\begin{split}
\mathcal{B}_1(\om) & =
\hat{a}_2(\om) \, \hat{\mathcal{R}}_{11}(\om)
- \hat{a}_1(\om) \,  \hat{\mathcal{R}}_{12}(\om) \cr
\mathcal{B}_2(\om) & =
\sum_{j=1}^{n}
\big(
\hat{a}_2(\om) - 2\hat{a}_1(\om) + \hat{a}_1(\om) \d_{i1} -
\hat{a}_{\frac{1}{2}}(\om) \d_{in}
\big)
\hat{\mathcal{R}}_{1i}(\om) \, .
\end{split}
\label{bulk_bound}
\ee
It is worth noting that explicit results on the generic boundary scattering amplitude
for the twisted Yangian are presented here for the first time (\ref{scatter}),
(\ref{bulk_bound}), although similar computations regarding the $\mathfrak{sl}(3)$ case were performed in \cite{Avan:2014noa}.
In the following we use the expressions on bulk scattering to explicitly show its factorization.

As in the $\frak{sl}(3)$ case \cite{Avan:2014noa}, it can be shown that the bulk scattering
factorizes to a product of a soliton$-$soliton times
a soliton$-$anti-soliton scattering amplitude of the usual Yangian
model. Consider two holes in the first sea, for the twisted Yangian
case. The bulk scattering phase in the
$\mathfrak{sl}({\cal N})$ (${\cal N} = 2n\ \mbox{or}\ 2n+1$)
twisted Yangian is given by
\be
{\cal B}_{\cal S}(\om) =
\hat{\mathcal{R}}_{11}(\om) \, \hat a_2(\omega)
- \hat{\mathcal{R}}_{12}(\om) \, \hat a_1(\om)
 =  {1-e^{\om} +
e^{-{{\cal N}\om \over 2}+ \om} -e^{-{{\cal N} \om \over 2}  }
\over 2 \sinh{{\cal N}\om \over 2} } \, .
\ee
In the usual $\frak{sl}({\cal N})$ case the soliton$-$soliton bulk
scattering amplitude is \cite{Doikou:1998xi}
\be
{\cal B}_{S}(\om) =
\frac{e^{-\om( {{\cal N} \over 2} - 1)}- e^{-{{\cal N} \om\over 2}}}{2\sinh ({{\cal N} \om\over 2})} \, ,
\ee
while for the soliton$-$anti-soliton we have \cite{Doikou:1998xi}
\be
{\cal B}_{\bar S}(\om) =
\frac{1-e^\om}{2\sinh ({{\cal N}\om \over 2})} \, .
\ee
The factorization of the scattering phase is then
immediately observed
\be
{\cal B}_{\cal S}(\om) = {\cal B}_{S}(\om) + {\cal B}_{\bar S}(\om) \, ,
\ee
leading to the factorization of the bulk part of the scattering
\be
\mathcal{S}_0(\l) = S_0(\l) \times \bar S_0 (\l) \, ,
\ee
where recall that $S_0,\ \bar S_0$ are the soliton$-$soliton and
soliton$-$anti-soliton scattering amplitudes in the Yangian
$\mathfrak{sl}({\cal N})$ case, expressed as:
\be
\mathcal{X}(\l)= \exp\Big[-\int_{-\infty}^{\infty}{d \omega \over \omega} e^{-i\om\lambda}\
{\cal B}_{\cal X}(\om)\Big],
~~~~~{\cal X} \in \{S,\ \bar S,\ {\cal S}\}.
\ee

We have shown that the factorization of the bulk
scattering, such as was first observed in the $\frak{sl}(3)$ twisted Yangian
case \cite{Avan:2014noa}, is valid in the generic case as well.  With this
we conclude our discussion on the bulk and boundary scattering in the
$\mathfrak{sl}({\cal N})$ twisted Yangian.

\section{Implementing defects}

We shall focus henceforth on the study of defect-transmission amplitudes in the
$\mathfrak{sl}({\cal N})$ twisted Yangian. As is well known, transmission
matrices physically describe the interaction between the
particle-like excitation of the model and the defect. It will be instructive in this section to introduce
some basic notions on the $\mathfrak{sl}({\cal N})$ twisted Yangian in the presence of defects.
The defect matrix used in this case takes of the generic form:
\be
{\mathbb L}(\lambda) = \lambda + i {\mathbb P} \, ,
\qquad {\mathbb P} = \sum_{i, j=1}^{\cal N} e_{ij} P_{ij} \, ,
\ee
$e_{ij}$ are ${\cal N} \times {\cal N}$ matrices such that
$(e_{ij})_{kl} = \delta_{ik}\, \delta_{jl}$,
and $P_{ij}$ are the $\mathfrak{gl}(\cal N)$ algebra generators
\be
\Big [P_{ij},\ P_{kl} \Big ] = \delta_{il} P_{kj} - \delta_{kj} P_{il} \, .
\ee
The Yangian $R$-matrix is given by
\cite{Yang:1967bm}
\be
R(\lambda) = \lambda + i{\mathcal P},
\qquad {\mathcal P} = \sum_{i, j=1}^{\cal N} e_{ij} \otimes e_{ji} \, , \label{ryang}
\ee
${\mathcal P}$ being the $\mathfrak{gl}({\cal N})$ permutation operator.
The $R$-matrix above is associated
with the fundamental representation of $\mathfrak{gl}({\cal N})$.

A generic finite-irreducible representation of the $\mathfrak{gl}({\cal N})$
algebra is associated with ${\cal N}$ integers
$\big(\a_1, \a_2, \ldots \a_{\cal N}\big)$, $\a_1 \geq \a_2 \geq \ldots \geq \a_{\cal N}$.
Here we deal with representations that possess highest weight state such that:
\ba
&& P_{kl}\, |\omega\rangle_n = 0, ~~~~k < l \cr
&& P_{kk}\, |\omega \rangle_n = \alpha_k\ |\omega \rangle_n \cr
&& e_{kl}\, | \omega \rangle_j = 0, ~~~~k>l, ~~~~j\neq n \cr
&& e_{kk}\, |\omega \rangle_j = \omega \rangle_j \,. \label{high}
\ea
The global reference state is then (note that we consider henceforth for convenience a chain with $L+1$ sites to incorporate the defect)
\be
|\Omega \rangle = \bigotimes_{j=1}^{L+1} | \om \rangle_j.
\ee

We shall also need the conjugate ${\mathbb L}$ matrix derived as
\be
\bar {\mathbb L}_{12}(\lambda) =
V_1\ {\mathbb L}^{t_1}_{12}(-\lambda -{i{\cal N}\over 2})\ V_1 \, , ~~~~~~~~V =\mbox{antidiag}(1, \ldots, 1)
\ee
where $^{t_1}$ denotes transposition on space 1, and hence
\be
\bar {\mathbb L}(\lambda) = \lambda + {i {\cal N}\over 2} - i \bar {\mathbb P} \, , ~~~~~~~~\bar {\mathbb P}_{12} = V_1\ {\mathbb P}_{12}^{t_1}\ V_1.
\ee
Note that physically the ${\mathbb L}$ matrix corresponds to the defect whereas the $\bar {\mathbb L}$ matrix correspond to what we figuratively call the anti-defect.

The transfer matrix of the twisted Yangian in the presence of defects reads as:
\ba
&& t(\lambda) = tr_0 \Big ({\mathbb K}_0^+(\lambda)\ {\mathrm T}_0(\lambda)\ {\mathbb K}_0^-(\lambda)\
V_0\ {\mathrm T}_0^{t_0}(-\lambda - {i{\cal N}\over 2})\ V_0 \Big ) \cr
&& {\mathrm T}_0(\lambda) = R_{0L+1}(\lambda) R_{0L}(\lambda) \ldots {\mathbb L}_{0n}(\lambda- \Theta) \ldots R_{01}(\l)
\ea
$R$ is the $\mathfrak{gl}({\cal N})$ Yangian matrix (\ref{ryang}), $\Theta$ is the rabidity associated to the defect.
Recall that we have considered for simplicity here, and throughout the text that
${\mathbb K}^{\pm} \propto {\mathbb I}$, and it is clear that the defect is inserted
in the $n^{th}$ site of the chain. It is also worth mentioning at this point that in order to obtain local Hamiltonians one needs to consider the alternating spin chain (alternate $R$ and $\bar R$ matrices) (see e.g.  \cite{Doikou:2000yw}, \cite{Arnaudon:2004sd}), the spectrum as well as the BAE are not modified in this case. The corresponding Hamiltonians in this case have been explicitly derived in \cite{Doikou:2000yw}, \cite{Arnaudon:2004sd} and contain terms that describe interactions up to four neighbours. Here of course we have to take into consideration the defect contributions which give rise to relevant terms that describe interactions up to six neighbours. Nevertheless, it is important to note that one can still obtain a local Hamiltonian via the usual process of taking the first derivative of the logarithm of the transfer matrix.

Assuming the existence of highest weight states which is the case here, the
formulation of the spectrum and Bethe ansatz equations follows the same logic
described in \cite{Doikou:2000yw}, \cite{Arnaudon:2004sd}, and in the case where defects are present
the BAE are modified accordingly as:
\begin{itemize}
\item $\mathfrak{sl}(2n+1)$
\ba
&& X_1^+(\lambda_i^{(1)} -\Theta)\ X_1^+(\lambda_i^{(1)} +\Theta)\ e^L_{1}(\l_i^{(1)}) =  \cr
&& \qquad -
\prod_{j=1}^{M^{(1)}} e_{2}(\l_i^{(1)} - \l_j^{(1)})\,
e_{2}(\l_{i}^{(1)} + \l_j^{(1)})
\prod_{j=1}^{M^{(2)}}
e_{-1}(\l_i^{(1)} - \l_j^{(2)}) \,
e_{-1}(\l_i^{(1)} + \l_j^{(2)})\,, \cr
&& X_{\ell}^+(\lambda_i^{(l)} -\Theta)\ X_{\ell}^+(\lambda_i^{(l)} +\Theta) =\cr
&& \qquad - \prod_{j=1}^{M^{(\ell)}}
e_{2}(\l_i^{(\ell)} - \l_j^{(\ell)}) \,
e_{2}(\l_i^{(\ell)} + \l_j^{(\ell)})
\prod_{ \tau = \pm1}
\prod_{ j=1}^{M^{(\ell+\tau)}} e_{-1}(\l_i^{(\ell)} - \l_j^{(\ell+\tau)}) \,
e_{-1}(\l_{i}^{(\ell)} + \l_{j}^{(\ell+\tau)}) \cr
&& \qquad \textrm{for} \qquad \ell = 2, \ldots, n-1, \cr
&& X_n^+(\lambda_i^{(n)} -\Theta)\ X_n^+(\lambda_i^{(n)} +\Theta)\ e_{-\frac{1}{2}}(\l_i^{(n)})
= - \prod_{ j=1}^{M^{(n-1)}}
e_{-1}(\l_i^{(n)} - \l_j^{(n-1)}) \,
e_{-1}(\l_i^{(n)} + \l_j^{(n-1)})  \,  \cr
&& \qquad \times - \prod_{j=1}^{M^{(n)}}
e_{-1}(\l_i^{(n)} - \l_j^{(n)})\,
e_{-1}(\l_i^{(n)} + \l_j^{(n)}) \,
e_2(\l_i^{(n)} - \l_j^{(n)}) \,
e_2(\l_i^{(n)} + \l_j^{(n)})
\ea
\item $\mathfrak{sl}(2n)$
\begin{eqnarray}
&& X_1^+(\lambda_i^{(1)} -\Theta)\ X_1^-(\lambda_i^{(1)} +\Theta)\ e_{1}^L(\l_{i}^{(1)}) = \cr
 &&  \qquad -\prod_{j=1}^{M^{(1)}}
  e_{2}(\l_{i}^{(1)} - \l_{j}^{(1)})\ e_{2}(\l_{i}^{(1)} +
  \l_{j}^{(1)})\ \prod_{ j=1}^{M^{(2)}}e_{-1}(\l_{i}^{(1)} -
  \l_{j}^{(2)})\ e_{-1}(\l_{i}^{(1)} + \l_{j}^{(2)})\,,
  \nonumber \\
 && X_{\ell}^+(\lambda_i^{(l)} -\Theta)\ X_{\ell}^-(\lambda_i^{(l)} +\Theta) =\cr
&& \qquad - \prod_{j=1}^{M^{(\ell)}} e_{2}(\l_{i}^{(\ell)} -
  \l_{j}^{(\ell)})\ e_{2}(\l_{i}^{(\ell)} + \l_{j}^{(\ell)})\
  \prod_{\tau = \pm 1}\prod_{ j=1}^{M^{(\ell+\tau)}}
  e_{-1}(\l_{i}^{(\ell)} -
  \l_{j}^{(\ell+\tau)})\ e_{-1}(\l_{i}^{(\ell)} +
  \l_{j}^{(\ell+\tau)}) \nonumber \\
  && \qquad \textrm{for} \quad \ell= 2,\ldots,n-1, \nonumber \\
 && X_n^+(\lambda_i^{(n)} -\Theta)\ X_n^-(\lambda_i^{(n)} +\Theta)\
e_{-1}(\l_{i}^{(n)}) \!\!=\!\! - \prod_{j=1}^{M^{(n)}}
  e_{2}(\l_{i}^{(n)} - \l_{j}^{(n)})\ e_{2}(\l_{i}^{(n)} +
  \l_{j}^{(n)}) \nonumber \\
  && \qquad \times \prod_{ j=1}^{M^{(n-1)}}e_{-1}^{2}(\l_{i}^{(n)} -
  \l_{j}^{(n-1)})\ e_{-1}^{2}(\l_{i}^{(n)} + \l_{j}^{(n-1)})\, ,
\label{BAE2}
\end{eqnarray}
where $X_k^{\pm}$ are the defect contributions (the action of ${\mathbb L}_{kk}$ and $\hat {\mathbb L}_{kk}$ on the local highest weight state (\ref{high})) and are defined as:
\be
X_k^+(\lambda) = {\l +i \alpha_k - {ik \over 2} \over \l +i \alpha_{k+1} - {ik\over 2} },
~~~~X_k^-(\lambda) = {\l -i \alpha_{{\cal N}- k+1} + {i({\cal N}-k) \over 2} \over \l -
i \alpha_{{\cal N}- k} + {i({\cal N}-k) \over 2} } \, .
\ee
\end{itemize}

Having derived the associated BAEs we now formulate a suitable
quantization condition for the model in the presence of defects. In order to
determine the relevant transmission matrix it suffices to consider a state with
one hole in the first sea. Before we discuss the quantization condition in this
case, let us first introduce some notation and define the
transmission amplitudes in $\mathfrak{gl}({\cal N})$ \cite{Doikou-trans} as
\ba
T(\lambda-\Theta):&& \mbox{soliton$-$defect scattering}\cr
\bar T(\lambda-\Theta):&& \mbox{soliton$-$anti-defect scattering}\cr
T^*(\lambda+\Theta):&& \mbox{anti-soliton$-$defect scattering}\cr
\bar T^*(\lambda+\Theta):&& \mbox{anti-soliton$-$anti-defect scattering}
\ea

The quantization condition for such a state reads as
\be
\Big (e^{i{\cal P}^{(l)}}{\mathbb S}(\tilde \l^{(l)}, \Theta) - 1\Big )
|\tilde \l^{(l)}, \Theta \rangle =0 \, ,
\ee
where the global scattering amplitude is given by
\be
{\mathbb S}(\lambda, \Theta) = {\cal K}^+(\lambda)\ T(\l - \Theta)\ \bar T(\l - \Theta)\
{\cal K}^-(\lambda)\  \bar T^*(\l +\Theta)\ T^*(\l +\Theta) \label{kt}
\ee
The latter can be pictorially represented as:

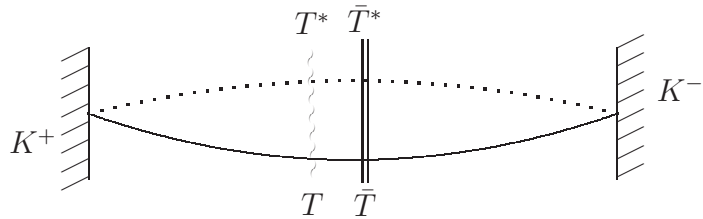
\begin{picture}(0,0)(-125,60)
\put(75,55){$\xymatrix{T^*\ar@{~}[dd] \cr \cr T}$}
\put(95,56){$\xymatrix{\bar{T}^*\ar@{=}[dd] \cr \cr \bar{T}}$}

\put(0,50){\line(0,-1){50}}
\multiput(0,50)(0,-7){8}{\line(-2,-1){10}}

\linethickness{0.1mm}
\qbezier(0,25)(100,-10)(200,25)

\linethickness{0.3mm}
\qbezier[40](0,25)(100,50)(200,25)

\linethickness{0.1mm}
\put(200,50){\line(0,-1){50}}
\multiput(200,50)(0,-7){8}{\line(2,1){10}}

\linethickness{0.7mm}
\put(-30,10){$K^+$}
\put(215,30){$K^-$}

\end{picture}

\vskip 2.7cm

The factorization of transmission amplitudes will be transparent in the following.
Note that the boundary scattering remains
unaffected and is given in the previous section.
The phase associated to the global defect$-$particle interaction in the
twisted Yangian is provided by
\be
{\cal B}_{\mathbb T}(\omega) = \sum_{k=1}^{n} \hat {\cal R}_{1k}(\om)
\Big ( \hat Y_k^{+}(\om) e^{i\om\Theta}+
\hat Y_k^{-}(\om) e^{-i\om\Theta}\Big ) \, ,
\ee
where we define
\be
Y^{\pm}_k(\l) = {i\over 2 \pi} {d X^{\pm}_k(\l) \over d \l} \, .
\ee
More precisely,
\ba
Y^{+}_k(\lambda) &=& {\mathrm a}(\a_k-{k\over 2},\ \a_{k+1} -{k\over 2};\l) \cr
Y^{-}_k(\lambda) &=& {\mathrm a}(-\a_{{\cal N}-k+1 }+ {{\cal N}- k\over 2},\ -\a_{{\cal N}-k} +{{\cal N}- k\over 2};\l)
\ea
and we define
\be
{\mathrm a}(x, y; \l) = {i\over 2 \pi} \Big ({1\over \l + ix} - {1\over \l + iy} \Big ) \, .
\ee
The Fourier transforms of the latter expressions are given below.
We distinguish three cases according to the values of $x,\ y$, and we end up with the following Fourier transforms in the
isotropic case (see also \cite{Doikou-trans}):
\ba
&& \hat {\mathrm a}(x,y;\omega)= e^{\omega x} ~~~\omega<0, ~~~~~\hat {\mathrm a}(x,y;\omega)= e^{\omega y}
~~~~\omega>0,~~~~~x>0,~~y<0\cr
&& \hat {\mathrm a}(x,y;\omega)= e^{\omega y} - e^{\omega x}~~~\omega > 0, ~~~~~
\hat {\mathrm a}(x,y;\omega)= 0 ~~~~\omega < 0,~~~~~x,\
y <0 \cr
&& \hat {\mathrm a}(x,y;\omega)= e^{\omega x} - e^{\omega y} ~~~\omega < 0, ~~~~~
\hat {\mathrm a}(x,y;\omega)= 0 ~~~~\omega > 0,~~~~~x,\
y > 0.
\ea
It is clear that in  the special case $x=-y = {n\over 2}$ the expressions above reduce to the familiar Fourier
transforms.

The important observation at this point here is that
\be
\hat {\cal R}_{1k}(\omega) = \hat R_{1k}(\om)+ \hat R_{{\cal N}-1 k}(\om) \, ,
\label{split}
\ee
where $\hat R$ is the Fourier transform of the inverse kernel in the
$\mathfrak{gl}(\cal N)$ Yangian \cite{Doikou:1998xi}
\be
\hat R_{ij}(\om) = e^{|\om|\over 2}
{ \sinh (\mbox{min}(i, j ){\om \over 2}) \sinh ({\cal N}-
\mbox{max}(i, j ){\om \over 2})\over \sinh{ \om \over 2}
\sinh ({{\cal N}\om\over 2})} \, .
\ee
The latter identity (\ref{split}) naturally leads to the factorization of the
transmission amplitudes as depicted diagrammatically in the quantization condition figure.
Indeed, as discussed in \cite{Doikou-trans}, the phases associated to transmission
amplitudes are derived as
\ba
&& {\cal B}_T(\om) = \sum_{k=1}^{n}\hat R_{1k}(\om) \, Y_k^+(\om) \, e^{i\om\Theta}\cr
&& {\cal B}_{\bar T}(\om) = \sum_{k=1}^{n}\hat R_{1k}(\om) \, Y_k^-(\om) \, e^{-i\om\Theta}\cr
&& {\cal B}_{T^*}(\om)= \sum_{k=1}^{n}\hat R_{{\cal N} -1 k}(\om) \, Y_k^+(\om) \, e^{i \om \Theta} \cr
&& {\cal B}_{\bar T^*}(\om) =\sum_{k=1}^{n}\hat R_{{\cal N} -1 k}(\om) \, Y_k^-(\om)\, e^{-i\om\Theta}\, .
\ea
Taking also into account that
\be
\hat R_{{\cal N}-1 k } = \hat R_{1 {\cal N} -k} \qquad
\mbox{and} \qquad Y_{k}^+(\l) = Y_{{\cal N}-k}^-(-\l) \, ,
\ee
we conclude that
\be
{\cal B}_{\mathbb T}(\om) = {\cal B}_T(\om) + {\cal B}_{\bar T}(\om)
+ {\cal B}_{T^*}(\om) + {\cal B}_{\bar T^*}(\om) \, .
\ee
The latter leads to the factorization of the defect$-$particle interaction as
described schematically in the figure and in equation (\ref{kt}). The boundary
scattering is separated, and as already mentioned is unaffected
by the presence of the defect. We identify the following quantities
\be
\mathcal{X}(\l)= \exp\Big[-\int_{-\infty}^{\infty}{d \omega \over \omega} e^{-i\om\lambda}\
{\cal B}_{\cal X}(\om)\Big],
~~~~~{\cal X} \in \{{\mathbb T},\ T,\ \bar T,\ T^*,\ \bar T^*\}.
\ee

This concludes our investigation on the scattering amplitudes in the $\mathfrak{sl}({\cal N})$ twisted Yangian.

\section{Discussion}

We investigate in the present article the generic scattering in the context of the $\mathfrak{sl}({\cal N})$
twisted Yangian. Our analysis is based on the solution of the Bethe ansatz
equations in the thermodynamic limit. In particular, in the thermodynamic limit
the ground state and low-lying excitations are identified. It is worth emphasizing that in the $\mathfrak{sl}(2n+1)$ case the Bethe ansatz equations are familiar and similar to the $\mathfrak{osp}(1|2n)$ case, whereas in the
$\mathfrak{sl}(2n)$ case they are not of any known form and are investigated here for the first time.

The scattering among the particle-like excitations is derived and as expected, turns out to be
factorized into a product of the
soliton$-$soliton times the soliton$-$anti-soliton scattering amplitudes of
the bulk $\mathfrak{sl}({\cal N})$ case. We also provide explicit expressions on the boundary scattering amplitudes
We have considered here the simplest
boundary matrices i.e. ${\mathbb K}^{\pm} \propto {\mathbb I}$. One of the key
points in this investigation together with the study of the boundary
scattering is the use of the suitable quantization condition
compatible with the underlying algebraic setting as well as the
corresponding physical interpretation. The quantization condition on the scattering derived in \cite{Avan:2014noa}
is clearly confirmed here by the fact
that the bulk scattering factorizes into the product of the soliton$-$soliton
and soliton$-$anti-soliton scattering amplitudes.

Furthermore, we consider the situation where a local integrable defect is inserted. This is achieved by introducing a generic representation of the underlying algebra in a particular site of the open spin chain.
Thus in addition to the bulk and boundary scattering we also investigate the scattering of particle-like excitations with the defect, and derive the associated transmission amplitudes. The key objects here again are the quantization condition together with the derivation of densities of the states in the thermodynamic limit. These lead to the identification of the global transmission amplitude, which turns out to factorize into a product of four distinct terms which describe the soliton$-$defect, soliton$-$anti-defect, anti-soliton$-$defect and anti-soliton$-$anti-defect interactions of the $\mathfrak{sl}({\cal N})$ spin chain \cite{Doikou-trans}.


\begin{thebibliography}{1}



\bibitem{fm}
L. Freidel, J. M. Maillet, Phys. Lett {\bf B262} (1991) 268.

\bibitem{cherednik}
I.V. Cherednik, Theor. Math. Phys. {\bf 61} (1984) 977.

\bibitem{sklyanin}
E.K. Sklyanin, J. Phys. {\bf A21} (1988) 2375.

\bibitem{Ols}
G.I. Olshanski
 ``Quantum Groups (1992)'',
Springer Lecture notes in Math. 1510

\bibitem{Avan:2014noa}
  J.~Avan, A.~Doikou and N.~Karaiskos,
  arXiv:1410.5991 [hep-th].

\bibitem{Doikou:2000yw}
  A.~Doikou,
J.\ Phys.\ A {\bf 33} (2000) 8797,
  [hep-th/0006197].

\bibitem{ragoucy}
V. Caudrelier, M. Mintchev, E. Ragoucy and P. Sorba, J.Phys. {\bf A38} (2005) 3431, [hep-th/0412159].

\bibitem{Arnaudon:2004sd}
  D.~Arnaudon, J.~Avan, N.~Crampe, A.~Doikou, L.~Frappat and E.~Ragoucy,
  J.\ Stat.\ Mech.\  {\bf 0408} (2004) P08005,
  [math-ph/0406021].

\bibitem{Arnaudon:2003zw}
  D.~Arnaudon, J.~Avan, N.~Crampe, A.~Doikou, L.~Frappat and E.~Ragoucy,
  Nucl.\ Phys.\ B {\bf 687} (2004) 257,
  [math-ph/0310042].

\bibitem{Korepin:1979hg}
  V.~E.~Korepin,
  Commun.\ Math.\ Phys.\  {\bf 76} (1980) 165.

\bibitem{Andrei:1983cb}
  N.~Andrei and C.~Destri,
  Nucl.\ Phys.\ B {\bf 231} (1984) 445.

  \bibitem{Grisaru:1994ru}
  M.~T.~Grisaru, L.~Mezincescu and R.~I.~Nepomechie,
  J.\ Phys.\ A {\bf 28} (1995) 1027,
  [hep-th/9407089].


\bibitem{Doikou:1998xi}
 B. Sutherland, Phys. Rev. B\textbf{12} (1975) 3795;\\
  A.~Doikou and R.~I.~Nepomechie,
  Nucl.\ Phys.\ B {\bf 521} (1998) 547
  [hep-th/9803118].



\bibitem{Yang:1967bm}
  C.~N.~Yang,
  Phys.\ Rev.\ Lett.\  {\bf 19} (1967) 1312.


\bibitem{Doikou-trans}
A. Doikou, JHEP 08 (2013) 103, aXiv:1304.5901 [hep-th].


\end{thebibliography}
\end{document}